\documentclass{article}

\usepackage{arxiv}

\usepackage[utf8]{inputenc} % allow utf-8 input
\usepackage[T1]{fontenc}    % use 8-bit T1 fonts
\usepackage{hyperref}       % hyperlinks
\usepackage{url}            % simple URL typesetting
\usepackage{booktabs}       % professional-quality tables
\usepackage{amsfonts}       % blackboard math symbols
\usepackage{nicefrac}       % compact symbols for 1/2, etc.
\usepackage{microtype}      % microtypography
\usepackage{lipsum}		% Can be removed after putting your text content
\usepackage{graphicx}
\usepackage{natbib}
\usepackage{doi}

\title{Unlocking the Potential of Medical Imaging with ChatGPT's Intelligent Diagnostics}

\author{\hspace{1mm}Ayyub Alzahem \\
	Robotics and Internet-of-Things Laboratory\\
	Prince Sultan University\\
	Riyadh, Saudi Arabia \\
	\texttt{aalzahem@psu.edu.sa} \\
	\And
	\hspace{1mm}Shahid Latif \\
	Robotics and Internet-of-Things Laboratory\\
	Prince Sultan University\\
	Riyadh, Saudi Arabia \\
	\texttt{slatif@psu.edu.sa} \\
        \And
	\hspace{1mm}Wadii Boulila \\
	Robotics and Internet-of-Things Laboratory\\
	Prince Sultan University\\
	Riyadh, Saudi Arabia \\
	\texttt{wboulila@psu.edu.sa} \\
        \And
	\hspace{1mm}Anis Koubaa \\
	Robotics and Internet-of-Things Laboratory\\
	Prince Sultan University\\
	Riyadh, Saudi Arabia \\
	\texttt{akoubaa@psu.edu.sa} \\
}

\begin{document}
\maketitle

\begin{abstract}
% \lipsum[1]
	Medical imaging is an essential tool for diagnosing various healthcare diseases and conditions. However, analyzing medical images is a complex and time-consuming task that requires expertise and experience. This article aims to design a decision support system to assist healthcare providers and patients in making decisions about diagnosing, treating, and managing health conditions. The proposed architecture contains three stages: 1) data collection and labeling, 2) model training, and 3) diagnosis report generation. The key idea is to train a deep learning model on a medical image dataset to extract four types of information: the type of image scan, the body part, the test image, and the results. This information is then fed into ChatGPT to generate automatic diagnostics. The proposed system has the potential to enhance decision-making, reduce costs, and improve the capabilities of healthcare providers. The efficacy of the proposed system is analyzed by conducting extensive experiments on a large medical image dataset. The experimental outcomes exhibited promising performance for automatic diagnosis through medical images.
\end{abstract}

% keywords can be removed

\section{Introduction}
% \lipsum[2]
% \lipsum[3]
\label{intrp}
Medical imaging diagnostics use various imaging techniques to help diagnose and treat medical conditions \cite{yadav2019deep, houssein2021deep, sarvamangala2022convolutional}. These imaging techniques range from simple X-rays to more complex methods such as magnetic resonance imaging (MRI), computed tomography (CT), nuclear medicine scans, and ultrasound. Medical imaging is essential for doctors and healthcare professionals to diagnose and treat various medical conditions. It allows for non-invasive visualization of internal organs, tissues, and bones, enabling doctors to identify potential problems and develop treatment plans \cite{sgouros2020radiopharmaceutical}. Medical imaging is utilized in various medical specialties, such as radiology, oncology, cardiology, neurology, and orthopedics. It can detect cancer, heart problems, brain abnormalities, and musculoskeletal injuries.

Advancements in medical imaging technology have considerably increased diagnostic accuracy and speed, enabling early identification and treatment of many medical disorders \cite{cui2021artificial, rehman2022novel}. As a result, medical imaging has evolved into a crucial component of modern healthcare, aiding in improving patient outcomes and saving lives. However, effectively interpreting medical imaging and establishing a diagnosis is difficult. Medical imaging interpretation is a complicated undertaking that requires specific training and skill. Radiologists, for example, must thoroughly grasp the imaging techniques employed and the anatomy and physiology of the examined areas. They must also be knowledgeable about the many forms of pathology and disease processes seen in medical imaging. As a result, medical image interpretation and diagnosis can be time-consuming and costly.

The motivation of this study is to address the limitations of existing autonomous diagnostic algorithms using medical images and to explore the potential of incorporating natural language processing (NLP) with deep learning (DL) techniques to improve the accuracy and efficiency of automated medical imaging diagnosis. The primary contributions of this research are:

\begin{itemize}
\item The proposal of a novel decision support system for generating automated diagnosis reports based on medical imaging data that incorporates DL and NLP techniques. The system is comprised of several components, including data collection and labeling, model training, label extraction, query generation, and ChatGPT API for NLP.
\item Investigation of the performance of the proposed system on a large dataset of medical images and demonstration of its promising performance for automatic diagnostics. The generated reports have been validated with an expert to ensure their accuracy and reliability.
\item Discussion of the potential benefits of incorporating ChatGPT into autonomous diagnostics using medical imaging, such as providing additional context and information to the algorithm, assisting in data augmentation, facilitating overcoming class imbalance, and identifying high-risk cases for urgent clinical attention.

\end{itemize}

The remainder of the article is structured as follows: Section 2 includes some of the most significant works on autonomous diagnoses with medical images. Section 3 discusses the proposed architecture in detail. Section 4 comprises experiments and outcomes. Finally, section 5 concludes with a brief summary.

\section{Related works}
\label{related}
Automatic disease diagnostics using medical imaging constitute an important tool to effectively solve many problems related to human error \cite{abdou2022literature, dhar2023challenges}. It enables doctors to promptly detect irregularities and potentially life-threatening disorders in a patient's body. Deep learning (ML/DL) approaches have considerably increased the accuracy and efficiency of medical imaging analysis, enabling more precise and efficient examination of X-rays, CT scans, and MRIs. This technology saves time and effort, reduces the risk of human error, and leads to more accurate diagnoses and improved patient outcomes. Early detection of diseases can also help with timely interventions, potentially saving lives and reducing healthcare costs. The following discusses some of the latest studies on automatic disease diagnostics with medical imaging.

Fernandes et al. \cite{fernandes2020automatic} discussed using deep neural networks (DNNs) in medical imaging diagnostics and proposed a novel algorithm to automatically generate compact DNN architectures for diagnostic support. The algorithm suggested in this study operates in two stages. First, during the deepening phase, the algorithm develops a deep neural network (DNN) by incorporating residual layers until the model starts overfitting the data. The second stage, the pruning phase, involves reducing the number of floating-point operations in the DNN model to create a DNN with a lower computational cost, guided by user preferences. This approach combines two distinct areas of DNN architecture exploration and pruning and has been evaluated on two medical imaging datasets with favorable outcomes.

Madani et al. \cite{madani2018deep} addressed the challenges of the high cost of annotation and the lack of accessible data in medical imaging by proposing a data-efficient DL classifier for cardiology prediction tasks. The proposed approach used pipeline-supervised models to focus on relevant structures, achieving high accuracies of 94.4% and 91.2% for echocardiographic view classification and binary left ventricular hypertrophy classification, respectively.

Mou et al. \cite{mou2021cs2} introduced a novel convolution neural network (CS-Net) capable of segmenting curvilinear structures from medical and biomedical images. The proposed CS-Net incorporates a self-attention mechanism in the encoder and decoder, allowing it to learn hierarchical representations of curvilinear structures. The performance of the proposed CS-Net was evaluated using 2D and 3D images from six different imaging modalities.

Lin et al. \cite{lin2021deep} developed DL classifiers utilizing deep networks, including VGG, ResNet, and DenseNet, for the automated diagnosis of metastasis in SPECT bone images. Their approach involves cropping the thoracic region and utilizing geometric transformations to augment the original data. The classifiers are fine-tuned to improve the performance of SPECT bone image diagnosis. The experimental results demonstrated the effectiveness of the proposed scheme for bone metastasis identification with SPECT imaging.

Lundervold et al. \cite{lundervold2019overview} provided a comprehensive review of the recent progress and challenges in ML for medical image processing, specifically emphasizing DL in MRI. The authors discussed the potential of DL for medical imaging technology, diagnostics, data analysis, and healthcare. Additionally, this review article highlighted significant references, open-source code, educational resources, data sources, and challenges associated with medical imaging.

Kassania et al. \cite{kassania2021automatic} proposed a DL-based framework for automatic COVID-19 classification in medical imaging, focusing on feature extraction. The researchers evaluated the performance of various popular deep convolutional neural networks, including DenseNet, Xception, VGGNet, InceptionV3, MobileNet, ResNet, NASNet, and InceptionResNetV2. The experimental findings depict that the DenseNet121 with a Bagging tree classifier achieved the highest classification accuracy of 99%.

Gao et al. \cite{gao2020feature} proposed a novel architecture called FT-MTL-Net for multi-task deep learning in medical image analysis. The approach combines the features of individual tasks during the early training stage to improve generalizability and computational efficiency. In addition, the FT-MTL-Net utilizes feature transfer by incorporating tasks from the same domain and domain views to enhance generalization. The authors evaluated the proposed approach on a Full Field Digital Mammogram dataset for breast cancer diagnosis. The results showed that the FT-MTL-Net outperformed competing models in classification and detection while achieving comparable results in segmentation.

Shi et al. \cite{shi2020review} reviewed the utilization of medical imaging, specifically X-ray and CT, in the battle against COVID-19 and how artificial intelligence (AI) can augment the capabilities of these imaging modalities. This comprehensive review encompasses the entire medical imaging and analysis pipeline involved in COVID-19, from image acquisition and segmentation to diagnosis and follow-up. Furthermore, the authors discuss how AI can automate the scanning process, enhance work efficiency by precisely delineating infections in images, and assist radiologists in making clinical decisions for disease diagnosis, tracking, and prognosis. Finally, the article highlights the latest medical imaging and radiology advancements in the fight against COVID-19, focusing on integrating AI with X-rays and CT scans.

Another study was conducted to improve the early detection of brain abnormalities through magnetic resonance imaging (MRI) \cite{talo2019convolutional}. The researchers utilized pre-trained deep learning models, such as Vgg-16, AlexNet, ResNet-18, ResNet-34, and ResNet-50, to classify MR images into normal and abnormal categories. The performance of these models was compared, with the ResNet-50 model achieving the highest accuracy of 95.23\% ± 0.6. The model may be a useful tool for clinicians in confirming their observations following the manual reading of MRI images and can be further evaluated with larger datasets of MRI images for brain abnormalities.

\subsection{Limitations of Existing Schemes and Motivation}

Existing AI-based automatic diagnostic algorithms using medical images have several shortcomings. Firstly, these algorithms are often trained on limited and biased datasets, leading to a lack of generalizability and performance degradation in real-world settings. Secondly, they can be affected by class imbalance, where there are significantly fewer samples in one class than in another. Thirdly, these algorithms often require significant amounts of labeled data for training, which can be costly and time-consuming to obtain. Fourthly, they may be unable to capture the nuances and complexities of medical imaging interpretation, leading to suboptimal diagnostic accuracy.

Incorporating ChatGPT into autonomous diagnostics using medical imaging can improve these algorithms' capabilities in various ways \cite{koubaa2023exploring,jeblick2022chatgpt}. Firstly, it can provide additional context and information to the algorithm, improving its ability to understand and interpret medical imaging. Secondly, ChatGPT can assist in data augmentation, generating additional labeled data to improve model performance. Thirdly, it can facilitate overcoming class imbalance by generating synthetic samples of underrepresented classes. Fourthly, high-risk cases for urgent clinical attention can be identified. ChatGPT can help to improve the accuracy and efficiency of automatic diagnostics with medical imaging.

\section{Proposed approach}
\label{related}

The proposed system uses deep learning and natural language generation techniques to generate high-quality diagnostic reports from medical images. The block diagram of the proposed system is shown in Fig.1. The main modules of the proposed architecture are briefly discussed in the following. 

%*********************Figure 1**********************
\begin{figure}[h!]
\centering
\includegraphics[page=1,width=0.8\textwidth]{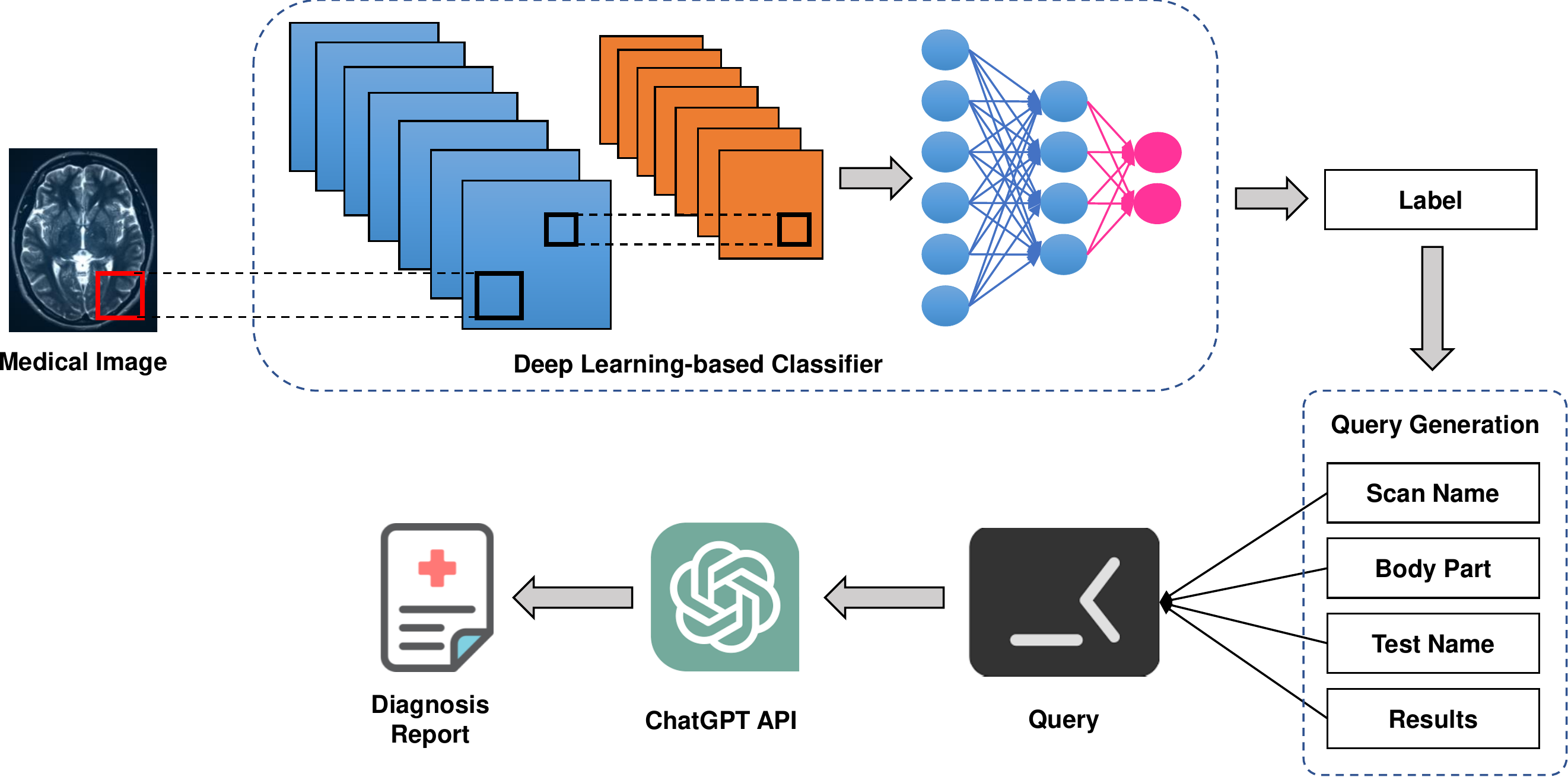}
\caption{Block diagram of the proposed architecture.}
\label{fig}
\end{figure}
%***************************************************

\subsection{Dataset Description}

\begin{figure}[h!]
\centering
\includegraphics[page=1,width=1.0\textwidth]{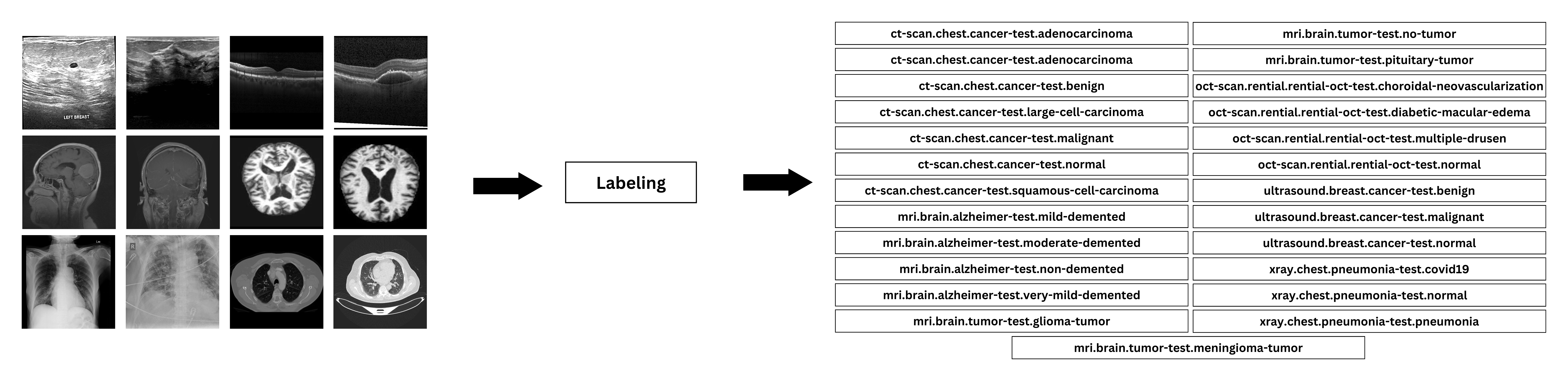}
\caption{Dataset collection and labeling.}
\label{fig:dataset}
\end{figure}

The utilized dataset contains a large collection of medical images manually labeled and split into training and validation sets. The dataset contains images from various medical imaging modalities, including MRI, CT, OCT, and ultrasound scans. In addition, the images are classified into different classes based on the medical condition or disease being examined. For example, the dataset includes images from different types of cancer, such as lung and breast cancer, and various stages of Alzheimer's disease, ranging from non-demented to moderately demented. It also includes images of different cancers, such as meningioma, pituitary, and glioma. Furthermore, the collection contains images of several respiratory disorders, such as pneumonia and TB, and OCT scans for diabetic macular edema, choroidal neovascularization, and numerous drusen. The data collection and labeling component is shown in Fig. \ref{fig:dataset} using our custom labels that contain numerous pieces of information for each label, allowing us to categorize the image based on the scan name, body part, test name, or disease. The dataset contains 152,856 images, including 122,257 in the training set and 30,599 in the validation set. The class distribution is varied, with some classes having a significant number of images, such as normal OCT scans, while others have relatively few images, such as benign breast cancer ultrasounds. Table \ref{tab:datasets} presents detailed quantitative information about the dataset. This dataset can be used to train and evaluate deep learning models for medical image classification problems.

%*********************Figure 2**********************
%***************************************************

\begin{table}[h]
\scriptsize
\centering
\caption{Quantitative distribution of dataset.}
\begin{tabular}{|l|r|r|}
\hline
\textbf{Label} & \textbf{Train} & \textbf{Val} \\ \hline
ct-scan.chest.cancer-test.adenocarcinoma & 1093 & 274 \\ \hline
ct-scan.chest.cancer-test.benign & 96 & 24 \\ \hline
ct-scan.chest.cancer-test.large-cell-carcinoma & 628 & 158 \\ \hline
ct-scan.chest.cancer-test.malignant & 448 & 113 \\ \hline
ct-scan.chest.cancer-test.normal & 1011 & 253 \\ \hline
ct-scan.chest.cancer-test.squamous-cell-carcinoma & 881 & 221 \\ \hline
mri.brain.alzheimer-test.mild-demented & 7884 & 1972 \\ \hline
mri.brain.alzheimer-test.moderate-demented & 5222 & 1306 \\ \hline
mri.brain.alzheimer-test.non-demented & 10242 & 2560 \\ \hline
mri.brain.alzheimer-test.very-mild-demented & 8960 & 2240 \\ \hline
mri.brain.tumor-test.glioma-tumor & 1881 & 471 \\ \hline
mri.brain.tumor-test.meningioma-tumor & 1316 & 329 \\ \hline
mri.brain.tumor-test.no-tumor & 400 & 100 \\ \hline
mri.brain.tumor-test.pituitary-tumor & 1464 & 367 \\ \hline
oct-scan.rential.rential-oct-test.choroidal-neovascularization & 29964 & 7491 \\ \hline
oct-scan.rential.rential-oct-test.diabetic-macular-edema & 9278 & 2320 \\ \hline
oct-scan.rential.rential-oct-test.multiple-drusen & 7092 & 1780 \\ \hline
oct-scan.rential.rential-oct-test.normal & 21254 & 5331 \\ \hline
ultrasound.breast.cancer-test.benign & 3780 & 945 \\ \hline
ultrasound.breast.cancer-test.malignant & 3553 & 889 \\ \hline
ultrasound.breast.cancer-test.normal & 106 & 27 \\ \hline
xray.chest.pneumonia-test.covid19 & 460 & 116 \\ \hline
xray.chest.pneumonia-test.normal & 1266 & 317 \\ \hline
xray.chest.pneumonia-test.pneumonia & 3418 & 855 \\ \hline
xray.chest.pneumonia-test.turberculosis & 560 & 140 \\ \hline
\end{tabular}
\label{tab:datasets}
\end{table}

\subsection{DL-based Classifier}
The second stage of the proposed architecture is a DL-based classifier that predicts the labels of provided medical images. DL algorithms can automatically learn complex features from raw image data, eliminating the need for hand-crafted feature extraction and reducing the time and effort required for preprocessing the data, making the overall process more efficient. DL approaches are well-suited to handling large datasets, which are common in medical imaging. They can learn from a large amount of data, improving the model's accuracy and generalization ability \cite{solano2020alzheimer, ben2022fusion}.

In the proposed scheme, we incorporated the DenseNet architecture for label prediction. DenseNet is a contemporary convolutional neural network (CNN) architecture for recognizing visual objects that has achieved cutting-edge performance with fewer parameters \cite{ben2022randomly, nandhini2022automatic}. DenseNet, with certain fundamental modifications, bears a striking resemblance to ResNet. However, in contrast to ResNet's additive attribute $(+)$ for merging previous and future layers, DenseNet utilizes a concatenated (.) attribute to merge the previous layer's output with that of the future layer. In addition, the issue of vanishing gradients can be resolved by the DenseNet Architecture, which connects all layers densely. In this study, we employed the DenseNet-121 architecture, which comprises 5 convolution and pooling layers, 3 transition layers (with sizes 6, 12, and 24), 1 classification layer (with a size of 16), and 2 dense blocks that employ $1 \times 1$ and $3 \times 3$ convolutions.

Typically, conventional CNNs generate output layers (lth) by applying a non-linear transformation $\psi_l($.$)$ to the output of the preceding layer $\mathcal{E}_{l-1}$.

\begin{equation}
\mathcal{E}l=\psi_l\left(\mathcal{E}{l-1}\right)
\end{equation}

DenseNets do not sum up the functionality maps of layer outputs with inputs but concatenate them. This communication model of DenseNet facilitates better information flow between layers, as each layer receives input from the feature maps of all preceding levels. This transformation can be expressed through the following equation:

\begin{equation}
\mathcal{E}_l=\psi_l\left[\left(\mathcal{E}_0, \mathcal{E}_1, \mathcal{E}2, \ldots, \mathcal{E}{l-1}\right)\right]
\end{equation}

The tensor $[\mathcal{E}_0, \mathcal{E}_1, \mathcal{E}2, \ldots, \mathcal{E}{l-1}]$ is formed by concatenating the output feature maps of previous layers, and the function $\psi_l(.)$ represents a non-linear transformation function, which includes three primary operations: batch normalization (BN), activation (ReLU), and convolution and pooling (CONV). Moreover, the growth rate $\beta$ plays a crucial role in determining the number of feature maps of the $l^{th}$ layer, as it is defined by the expression $\beta^{[l]}=\beta^{[0]}+\beta(l-1)$, where $\beta^{[0]}$ denotes the number of feature maps in the first layer.

\subsection{Query Generation}

The Query Generation component of our system is responsible for generating a prompt that can be used as input to the ChatGPT API. The prompt describes the medical imaging data, including the scan name, body part, test name, and test result, and asks the user to write a medical report. The prompt also includes a section for relevant prescriptions and possible causes of detected diseases. Our Query Generation function uses the label from the medical imaging dataset as input and extracts the relevant information to generate a coherent prompt.

\subsection{ChatGPT API}

The ChatGPT API is a natural language processing tool created by OpenAI that may be used for various activities such as text completion, summarization, and translation. In our system, we utilize the ChatGPT API to produce a medical report based on the prompt given by the Query Generation component. The designed API is used to create a response that contains a description of the medical imaging data, appropriate medications, and likely causes of diagnosed disorders. The proposed solution uses the "gpt-3.5-turbo" version of the ChatGPT API, which has quicker response times and higher accuracy than prior versions. The suggested system illustrates how the ChatGPT API may create automated medical reports based on medical imaging data, possibly enhancing diagnostic efficiency and accuracy \cite{koubaa2023gpt}.

\subsection{Diagnosis Report}

Our system provides a diagnosis report summarizing the medical imaging data, including any important diagnoses and prescriptions. The ChatGPT API creates the report depending on the prompt provided by the Query Generation module. The report may include information such as the type of scan performed, the body portion investigated, and any abnormalities discovered. Furthermore, the report may contain potential causes of diagnosed conditions and proposed medical treatments. Moreover, our system provides the diagnosis report promptly, allowing doctors to make more accurate decisions. Finally, the suggested approach is a potential step toward employing natural language processing techniques to automate medical diagnosis.

\section{Experiments}

\subsection{Implementation Platform}

The proposed model is trained, and performance is investigated on an Intel Core i9-9900K CPU @ 3.6 GHz. The system is equipped with 64GB RAM and a 12GB NVIDIA GeForce RTX 2080 graphics card to ensure the smooth execution of DL algorithms. The proposed model is simulated in Google Colab, and the Jupyter notebook of the proposed implementation can be provided on request for future endeavors.

\subsection{Model training}

Model training is an essential process in machine learning that involves learning patterns and connections between input data and output labels. In this case, we trained several pre-trained models on the given dataset, and the Densenet121 model achieved the highest accuracy with a batch size of 16, a learning rate of 0.0001, and 100 epochs. Tuning the batch size and learning rate is crucial in model training, as they determine how the model updates its parameters during training. A smaller batch size may lead to noisy updates, while a larger one may take longer to converge. Similarly, a larger learning rate might push the model beyond its ideal parameters, while a lower learning rate could lead to slower convergence. Therefore, hyperparameters should be carefully tuned to obtain the best results.

\begin{figure}[h!]
\centering
\includegraphics[page=1,width=0.8\textwidth]{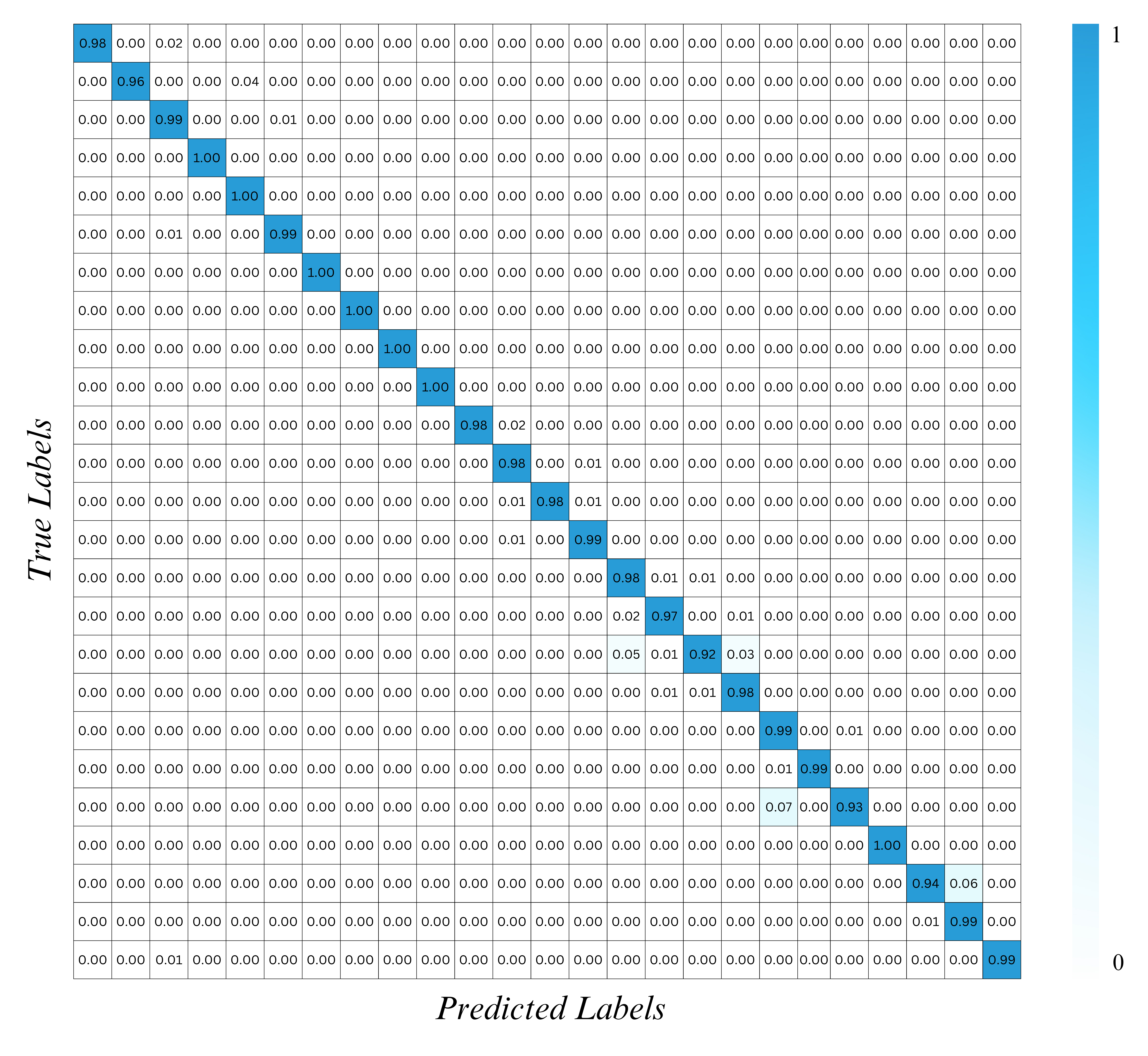}
\caption{DenseNet121 confusion matrix on our medical dataset.}
\label{fig:cm}
\end{figure}

The Densenet121 model achieved a training loss of 0.0002 and a training accuracy of 0.9984, showing that the model has figured out the correlations and patterns between the input characteristics and the output labels. We then used a validation set to assess how well the model performs with unknown data. The model achieved a validation loss of 0.0059, a validation accuracy of 0.9820, a precision of 0.9821, a recall of 0.9820, and an F1 score of 0.9820. These measures assess the performance of the model on the validation set, where the validation loss represents the discrepancy between expected and actual labels, and the validation accuracy indicates the proportion of correctly predicted output labels. The precision counts the percentage of accurate positive forecasts among all positive predictions, whereas the recall counts the percentage of accurate positive cases. Finally, the F1 score is a balanced metric that takes accuracy and recall into account and is the harmonic mean of these two metrics. Figure \ref{fig:cm} presents the confusion matrix of our model.

%*********************Figure**********************
%***************************************************

To train the model, we used the Adam optimizer, an optimization algorithm that updates the model parameters during training using adaptive learning rates, leading to faster convergence and better performance. We also used the categorical cross-entropy loss function, a typical loss function used in classification problems, to minimize the difference between expected and actual output labels.

Table \ref{tab:Hyperparameters} summarize the hyperparameters  we used in the model training.

\begin{table}[h]
\scriptsize
\centering
\caption{The utilized hyperparameters.}
\begin{tabular}{|c|c|}
\hline
\textbf{Model Name} & DenseNet121 \\ \hline
\textbf{Optimizer} & Adam \\ \hline
\textbf{Loss Function} & Cross-Entropy \\ \hline
\textbf{Batch Size} & 16 \\ \hline
\textbf{Learning Rate} & 0.0001 \\ \hline
\textbf{Epochs} & 100 \\ \hline
\end{tabular}
\label{tab:Hyperparameters}
\end{table}

\subsection{Reports Analysis}
In our experiments, we generated 25 reports using the proposed framework and different medical images. These reports summarize medical images, diagnosed diseases, and prescriptions. In the following section, we discuss two diagnostic reports.

Alzheimer's disease is a progressive neurological disorder that impacts memory and cognitive function. It is the leading cause of dementia in the elderly, and its incidence is predicted to rise as the population ages. Alzheimer's disease is frequently diagnosed using a combination of medical history, physical examination, and cognitive testing. Medical imaging, such as MRI scans, can also aid in identifying diseases and monitoring their progression.

The first report featured an MRI scan image of a patient with Alzheimer's disease. According to the research findings, the patient was at a moderate stage of the condition, suggesting considerable cognitive decline and functional disability. The report also pointed out various potential causes of Alzheimer's disease, including genetic factors, brain damage, and lifestyle factors such as smoking, high blood pressure, and high cholesterol. These risk factors can raise the probability of getting the condition and influence its development. Therefore, the proposed system recommended several precautions and treatment strategies for the patient. First, the report advised the patient to consult with a neurologist and follow the indicated medication. Treatment options may include a combination of medications and lifestyle changes such as good nutrition, exercise, and cognitive therapy. These therapies can help prevent the disease's progression and enhance the patient's quality of life. The report additionally recommended regular follow-up sessions to monitor the patient's condition and ensure a successful treatment plan. These sessions also provide an opportunity to discuss any issues or queries the patient or their family may have.

%*********************Figure 3**********************
\begin{figure}[h]
\centering
\includegraphics[scale=0.33]{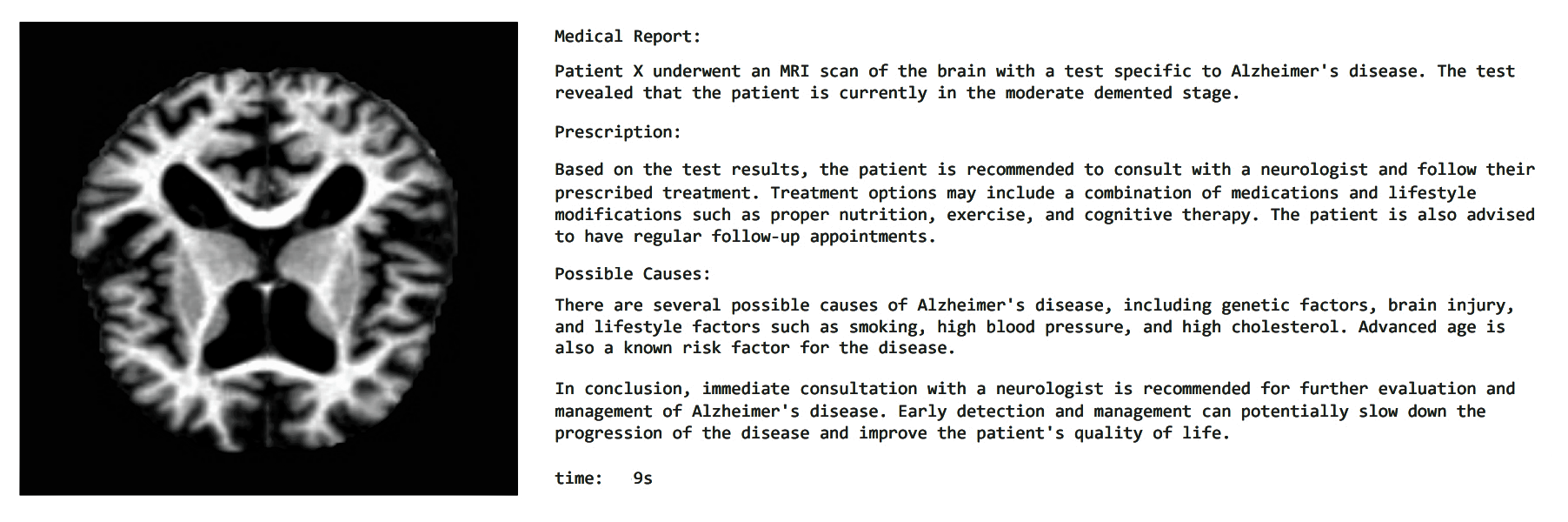}
\caption{Experimental results of Alzheimer's disease from an MRI image.}
\label{fig:report1}
\end{figure}
%***************************************************

\begin{figure}[h]
\centering
\includegraphics[scale=0.25]{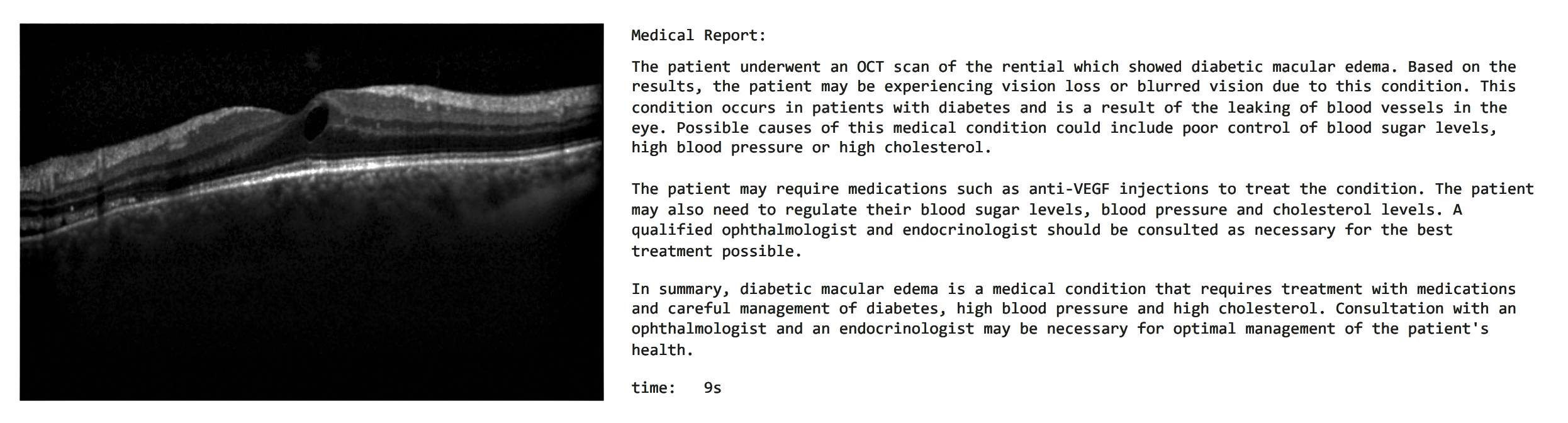}
\caption{Experimental results of diabetic macular edema from an OCT scan image.}
\label{fig:report2}
\end{figure}

The second report in Fig. \ref{fig:report2} summarizes the findings of an experiment on diabetic macular edema, a frequent consequence of diabetes. An OCT scan of the retina was performed on the patient, confirming the presence of diabetic macular edema. When blood vessels in the eye start to leak, fluid accumulates in the macula, which is the central part of the retina crucial for clear vision. As a result of this disease, the patient may experience vision loss or blurred vision, which can have a negative impact on their quality of life.

Diabetic macular edema can be caused by various factors, including poor blood sugar management, high blood pressure, or high cholesterol. These conditions can cause damage to blood vessels in the eye, leading to fluid leakage and enlargement of the macula. To treat diabetic macular edema, the patient may require drugs such as anti-VEGF injections to reduce swelling and improve vision. To prevent further damage to the blood vessels in the eye, the patient may also need to maintain their blood sugar, blood pressure, and cholesterol levels, in addition to medication. Therefore, it is crucial to consult with a trained ophthalmologist and endocrinologist to develop a personalized treatment plan based on the patient's individual needs. By effectively managing their disease, the patient can reduce their risk of vision loss while maintaining their overall health and well-being.

%*********************Figure 4**********************
%***************************************************

The proposed scheme aims to revolutionize the current medical report generation process by offering a faster and more efficient diagnostic technique. The suggested framework enables clinicians to quickly produce a comprehensive diagnostic report, providing them with more time to make informed decisions about the patient's condition. The primary advantage of this approach is its utilization of the latest developments in NLP, which improves diagnostic accuracy while minimizing the chances of misinterpretation. The NLP capabilities of our system evaluate medical imaging data and transform it into a readable format that medical practitioners can easily understand. The solution reduces human errors and eliminates the need for time-consuming manual labor by automating the diagnostic procedure. The system is a promising step toward automating medical diagnosis since it can assist medical practitioners in making more accurate diagnoses in less time. Additionally, the proposed framework ensures that patients receive timely treatment, which may lead to better outcomes. By creating precise and efficient diagnostic reports based on medical imaging data, the suggested system enables clinicians to take prompt action. This approach has the potential to significantly improve the efficiency and accuracy of the diagnostic process, resulting in better patient outcomes. This method is especially valuable in emergency situations, where a swift and accurate diagnosis is critical to a patient's survival. The proposed system can be embedded in service-oriented monitoring, diagnostics, and control toward better healthcare decision support \cite{chandrasekar2022lung,driss2020servicing}.

\section{Conclusion}

This article proposes a decision support system that uses DL and NLP techniques to generate automated diagnosis reports based on medical imaging data. The system contains several components: data collection and labeling, model training, label extraction, query generation, and the ChatGPT API for NLP. The system's performance was investigated on a large dataset of medical images, and the results demonstrated promising performance for automatic diagnostics. One of the system's main advantages is that it provides a faster and more efficient way of generating medical reports compared to traditional methods. In the proposed framework, the utilized DenseNet121 model achieved the highest accuracy of 98\%. The proposed system represents a promising solution for generating high-quality medical reports from medical images. Future perspectives of this work involve assessing the performance of ChatGPT-4 and the possibility of integrating it into a heart care diagnosis system.

\bibliographystyle{unsrtnat}
\bibliography{references}  %%% Uncomment this line and comment out the ``thebibliography'' section below to use the external .bib file (using bibtex) .





% \end{thebibliography}

\end{document}